\documentclass[preprint]{article}

\usepackage{amsmath}
\usepackage{commath}
\usepackage[hidelinks, hypertexnames=false]{hyperref}
\usepackage{graphicx}
\usepackage{siunitx}

\newcommand{\D}{\mathrm{d}} 		
\newcommand{\E}{\mathrm{e}} 		
\newcommand{\I}{\mathrm{j}} 			
\newcommand{\J}{\mathrm{j}} 			

\usepackage{amssymb}

\title{Difference between charge--voltage relations of ordinary and fractional capacitors}

\author{Eirik Brenner Marthins, Sverre Holm\\
Physics Department, University of Oslo, Oslo, Norway}

\begin{document}

\maketitle

\begin{abstract}
In an ordinary time-varying capacitor, there is debate whether a time-domain multiplication or a time-domain convolution of capacitance and voltage determines charge. A time-varying capacitor in parallel with a resistor was implemented by a motor-driven potentiometer and op-amps. The response matched a power-law function over about two decades of time, and not an exponential, for several sets of parameters. This confirms the time-domain multiplication model. This result is the opposite of that obtained for a Constant Phase Element (CPE) in its common time- and frequency-varying capacitor interpretation. This demonstrates that a CPE is fundamentally different from an ordinary time- and frequency-varying capacitor.  
\end{abstract}


\section{Introduction}
In an ideal capacitor with capacitance $C$, the charge -voltage relationship in the time and frequency domains are:
\begin{equation}
    q(t)=C\cdot v(t) \enspace \Leftrightarrow \enspace Q(\omega)=C\cdot V(\omega),
    \label{eq:Q-constantC}
\end{equation}
where lower-case letters denote the time-domain, and capital letters denote the Fourier domain.

The simple relation of \eqref{eq:Q-constantC} is no longer true when the capacitance varies with time as pointed out in \cite{allagui2021revisiting}. As a multiplication in one domain is equivalent to a convolution in the other, only one of the following alternatives can be valid. The first one is:
\begin{equation}
    q(t)=c(t)\cdot v(t) \enspace \Leftrightarrow \enspace Q(\omega)=C(\omega)\ast V(\omega).
    \label{eq:Q-multiplicationTime}
\end{equation}
This time-domain multiplication is assumed in  \cite{holm2021simple} and recommended in \cite{jeltsema2022time, ortigueira2023new}.

The alternative, based on time-domain convolution, is 
\begin{equation}
    q(t)=c(t)\ast v(t)\enspace \Leftrightarrow \enspace Q(\omega)=C(\omega)\cdot V(\omega),
    \label{eq:Q-convolutionTime}
\end{equation}
which is recommended in  \cite{allagui2021revisiting, allagui2023tikhonov}  based on analysis of and measurements on  systems with fractional capacitors, also called constant phase elements (CPE).

Here we report results from measurements on a time-varying non-fractional capacitor which lend support to the expression with time domain multiplication, \eqref{eq:Q-multiplicationTime}. We then show that ordinary and fractional capacitors will follow different rules. This explains why different and seemingly conflicting results have been reported in the recent literature. 

We first recall theory for time-varying capacitors and find the consequences of the two alternative models. We then analyze CPEs showing its charge--voltage response and  the relationship with classical dielectric models, the Curie-von Schweidler current response, and the Kohlrausch charge response. Then we discuss how an ordinary time-varying  capacitor with enough range of variation can be built using op-amps and a motor-driven potentiometer. Finally we report measurements that distinguish between the two alternatives.

\section{Theory}

The current-charge relation is by definition: 
\begin{equation}
    i(t)=\frac{\D }{\D t} q(t).
    \label{eq:Ccurrent}
\end{equation}
This means that there are also two alternative  voltage-current relations corresponding to \eqref{eq:Q-multiplicationTime} and \eqref{eq:Q-convolutionTime}:
\begin{equation}
i(t)=c(t) \frac{\D v(t)}{\D t}+\frac{\D c(t)}{\D t}v(t),
\label{eq:I-TimeMultiplication}
\end{equation}
or
\begin{equation}
i(t)=\frac{\D}{\D t} \left[ c(t)\ast v(t) \right].
\label{eq:I-TimeConvolution}
\end{equation}
\subsection {Temporal responses of linearly increasing capacitance}
\label{sec:TempLinIncrC}
\begin{figure}[tb]
\centering
\includegraphics[scale=0.15]{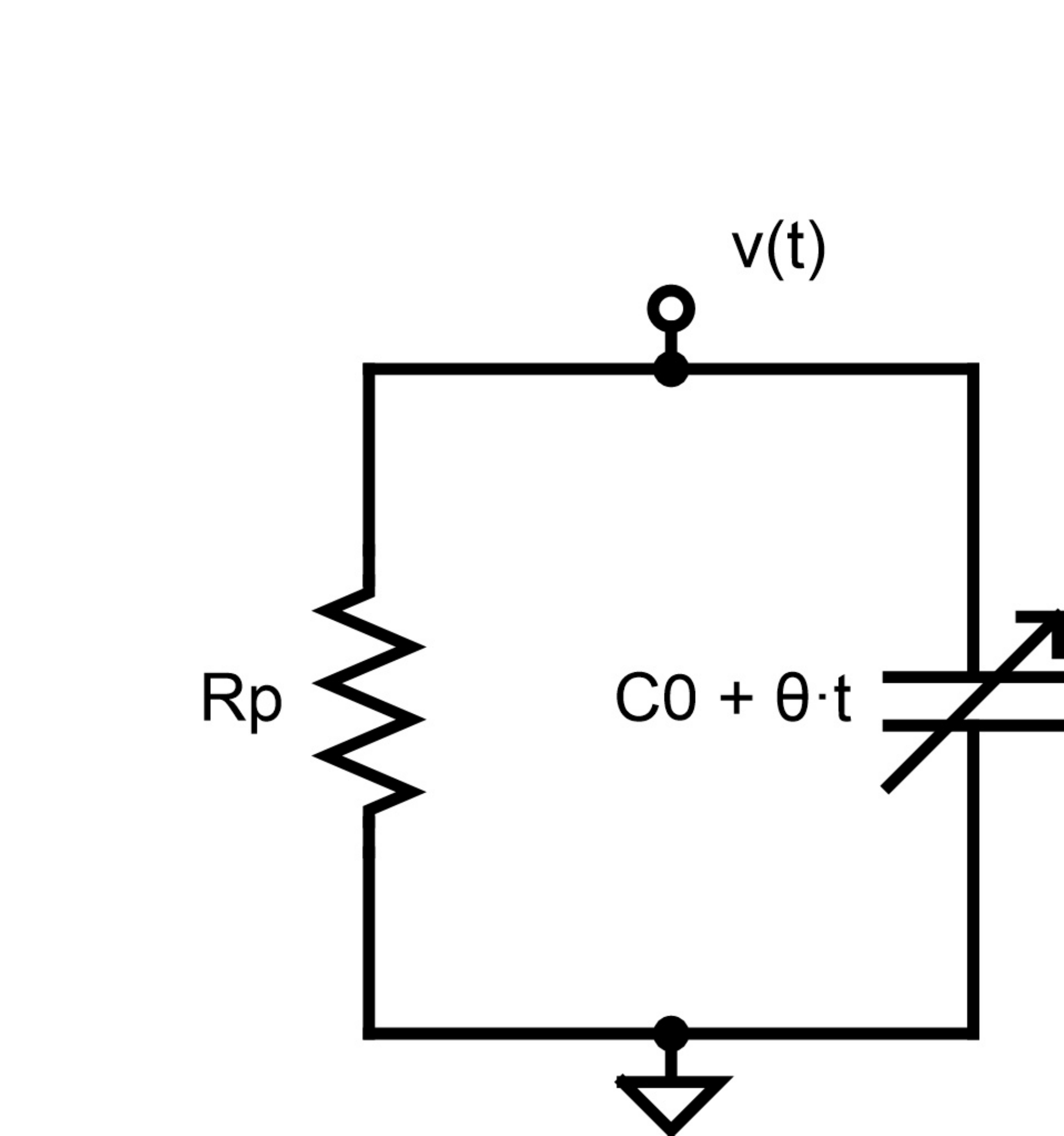}
\caption{Time-varying capacitor in parallel with resistor (Drawn using \url{https://www.circuit-diagram.org/}).}
\label{fig:RVariableC}
\end{figure}
The capacitor is assumed to increase linearly with time
\begin{equation}
    C(t')=C_0+\theta t'.
    \label{eq:TimeVaryingC}
\end{equation}
As is usual in time-varying systems \cite{pandey2016linking}, the time variable, $t'$, for when the value starts increasing is in general independent of the time of application of an excitation, $t$. Often the two are the same, as in the following derivation.

The capacitor is connected in parallel with a resistor $R_p$ as shown in Fig.~\ref{fig:RVariableC}.
 Assume  that the inital value for the voltage is $v(0)=V_0$. In the following we will find what is both the particular (forced) solution resulting from an input current impulse that results in $V_0$, and the homogeneous (unforced) solution from an initial condition. The latter case is what we will measure later. 
 
 It is first assumed that the multiplication relation of \eqref{eq:I-TimeMultiplication} is valid. In both the forced and the unforced cases, the circuit of Fig.~\ref{fig:RVariableC} will have a voltage response which according to \cite{holm2021simple} and \cite[Appendix]{pandey2016linking} is:
\begin{equation}\label{eq:LCPEimpulse2}
    v(t)=V_0 \left(1+\frac{t}{\tau}\right)^{-\alpha-1},  \,  \tau = \frac{C_0}{\theta}, \;  \alpha=\frac{1}{R_p\theta}, \; t>0.
\end{equation}
In the limit as $t\gg \tau$, this is a power-law function where the voltage follows
\begin{equation}
v(t) \approx V_0 (t/\tau)^{-\alpha-1}.    
\label{eq:ApproxLinearlyVaryingC}
\end{equation}

A simplified version of the multiplication relation is assumed in some circuit simulation tools where only the first term in \eqref{eq:I-TimeMultiplication} is included \cite{holm2021simple}. This model will also give a power-law result but it will vary as $(t/\tau)^{-\alpha}$. Thus it is easy to distinguish from \eqref{eq:ApproxLinearlyVaryingC}.

Finally follows an analysis of the response if the convolution relation of \eqref{eq:I-TimeConvolution} is assumed. First, it has to be realized that the time-varying capacitance expression in \eqref{eq:TimeVaryingC} implicitly is in the frequency domain. Therefore a constant capacitor, $C(\omega)=C_0$, corresponds to $c(t)=C_0\delta(t)$ in the time domain, where $c(t)$ has unit F/s. Likewise a linearly increasing capacitor is the integral of a constant capacitor, and the integral of the delta function $\delta(t)$ is the step function $u(t)$, giving
\begin{equation}
    c(t)=C_0\delta(t)+\theta u(t).
\end{equation}
It may be noted that this result is the temporal derivative of \eqref{eq:TimeVaryingC} when an implicit step function is assumed in \eqref{eq:TimeVaryingC}. This may explain why the controversial additional differentation appears in Eq.~(9) of \cite{pandey2022origin}, see discussion in \cite{allagui2022comment, jeltsema2022time}.

Inserted in \eqref{eq:I-TimeConvolution}, the current is:
\begin{align}\nonumber
    i(t) &=\frac{\D}{\D t} \left[C_0\delta(t)\ast v(t)+\theta u(t)\ast v(t)\right] +  {v(t)}/{R_p}\\
    &=\frac{\D}{\D t} \left[C_0v(t)+\theta  \int v(t) \D t\right]+v(t)/R_p
    =C_0\frac{\D v(t)}{\D t} +\left( \theta + 1/R_p \right) v(t).
\end{align}
The voltage response to a current impulse is found by setting the above expression to 0 and this gives a differential equation with solution
\begin{equation}
    v(t) = V_0 \E^{-t/\tau'}, \enspace \tau' =\frac{C_0}{\theta + 1/R_p},
    \label{eq:Exponential}
\end{equation}
where $V_0$ is an initial value determined by the amplitude and duration of the input current impulse.  

Thus, in order to distinguish whether \eqref{eq:I-TimeMultiplication} or \eqref{eq:I-TimeConvolution} is valid for a time-varying capacitor, one needs to find the voltage response to a current impulse  or an inital voltage and check whether it follows the power-law response of \eqref{eq:ApproxLinearlyVaryingC} or the exponential response of \eqref{eq:Exponential}. If it follows a power law, it should be checked whether the exponent is $-\alpha-1$ or just $-\alpha$.

\subsection{Constant phase element}
The constant phase element (CPE) is a circuit element with a history going back to Cole \cite{cole1928electric, cole1940permeability}.  It is defined by the impedance in the frequency domain, \cite{westerlund1994capacitor}:
\begin{equation}
{Z}(\omega) =\frac{{V}(\omega)}{{I}(\omega)}= \frac{1}{(\I \omega)^{\alpha'} C_{\alpha'}}, \enspace 0<\alpha' \le 1.
\label{eq:fractionalC}
\end{equation}
The equivalent capacitance is: 
\begin{equation}
    \tilde{C}(\omega) = (\I \omega)^{\alpha'-1} C_{\alpha'}.
\label{eq:EquivalentCFractional}
\end{equation}
As  $\alpha'$ approaches unity, it becomes an ordinary capacitor and for $\alpha'=0$ it simplifies to an ordinary resistor.
Inverse Fourier transformation of \eqref{eq:EquivalentCFractional} gives:
\begin{equation}
    \tilde{c}(t) = \frac{C_{\alpha'}}{\Gamma(1-\alpha')} t^{-\alpha'}
    \label{eq:EquivalentCFractionalTime}
\end{equation}
On purpose a tilde is used for  $\tilde{C}(\omega)$ and $\tilde{c}(t)$ as these are not capacitances in the same sense as $C(\omega)$ and $c(t)$ in Sect.~\ref{sec:TempLinIncrC}.

In the time domain, the current-voltage relation is given by a fractional derivative: 
\begin{equation}
i(t) = C_{\alpha'} \frac{\D^{\alpha'} v(t)}{\D t^{\alpha'}}.
\label{eq:fractionalDerivC}
\end{equation}

\subsubsection{Charge-voltage relation for CPE}
When \eqref{eq:Ccurrent} is combined with \eqref{eq:fractionalDerivC}, the result is
\begin{equation}
    \frac{\D^{1-\alpha'}}{\D t^{1-\alpha'}} q(t)=C_{\alpha'} \,v(t).
    \label{eq:Q-Fractional}
\end{equation}
This is a result from \cite{ortigueira2023new} where it is also shown that in the case of a time-varying CPE, the relation is
\begin{equation}
    \frac{\D^{1-\alpha'}}{\D t^{1-\alpha'}} q(t)=C_{\alpha'}(t)\,v(t).
    \label{eq:Q-TimeVaryingFractional}
\end{equation}
On the one hand, for the special case $\alpha'=1$ where the model becomes an ordinary capacitor, \eqref{eq:Q-TimeVaryingFractional}  will simplify to \eqref{eq:I-TimeMultiplication}, and  \eqref{eq:Q-Fractional} to the time-domain multiplication of \eqref{eq:Q-multiplicationTime}.

On the other hand, when \eqref{eq:Q-Fractional} is solved for the charge, the result is:
\begin{equation}
    q(t) = C_{\alpha'} \frac{\D^{\alpha'-1}}{\D t^{\alpha'-1}} v(t) = \tilde{c}(t) \ast v(t).
    \label{eq:Q-Fractional2}
\end{equation}
This is the convolution description of \eqref{eq:Q-convolutionTime} and the result is a consequence of the frequency domain definition of the CPE in \eqref{eq:fractionalC}. 

With the ordinary time-varying  capacitor, there may be uncertainty as to which of the charge-voltage responses are valid as shown in the previous section. Therefore experimental evidence, as in the remainder of this paper, is needed to decide between the two models. This is different for the CPE as the charge--voltage response is given by a convolution in time due to the way it has been defined in \eqref{eq:fractionalC}. 

\subsubsection{Temporal response of CPE}

Due to the singularity as the frequency approaches zero, it is evident that the CPE is an idealized circuit element. As capacitance is $C = \varepsilon_0 \varepsilon_r A/d$,
where $A$ is plate area, $d$ is plate distance, and $\varepsilon_0$ is the permittivity of vacuum, the CPE also corresponds to a singular model for relative permittivity:
\begin{equation}
    \varepsilon_r(\omega) =  \frac{\varepsilon_s }{(\I \omega \tau_C)^{1-\alpha'}},
    \label{eq:CPEpermitivity}
\end{equation}
where $\tau_C$ is a characteristic relaxation time, and $\varepsilon_s =  C_\alpha' \tau_C^{1-\alpha'}$. 

It is not uncommon to inverse transform \eqref{eq:CPEpermitivity} and get the Curie--von Schweidler law found below, see e.g.~\cite{das2017revisiting, pandey2022origin, ortigueira2023new}. But as this law also is singular, we find that it gives more insight to view both the CPE and the Curie-von Schweidler law as limiting cases of the response of well-behaved standard dielectric models, \cite{holm2020time}.

The most general of these dielectric models, the Havriliak--Negami model, has a relative permittivity of
\begin{equation}
{\varepsilon_r}(\omega) = 
\varepsilon_\infty + \frac{\varepsilon_s - \varepsilon_\infty}{\left( 1 +( \I \omega \tau_C)^{\gamma} \right)^\beta}, \enspace 0 \le \gamma \le 1, \; 0 \le \beta \le 1.
\label{eq:HN}
\end{equation}
Assuming $\varepsilon_\infty = 0$, $\omega \tau_C \gg 1$, and $\gamma \cdot \beta = 1-\alpha'$, will lead to \eqref{eq:CPEpermitivity}. Therefore the CPE can be found in the limit from either the Havriliak--Negami model, the Cole--Davidson model ($\gamma=1$), or the Cole--Cole model ($\beta=1$).

Here the latter case will be analyzed. In \cite{holm2020time} it is shown that the time-domain current step response is proportional to the response function, $\phi(t)$, of \cite{garrappa2016models}. The response function is the inverse Fourier transform of the relative permittivity in normalized form, $\varepsilon' (\omega)= \left(\varepsilon_r(\omega)- \varepsilon_\infty\right)/\left(\varepsilon_s - \varepsilon_\infty \right)$. For the Cole--Cole model, it is:
\begin{equation}
 \phi(t) =  \frac{1}{\tau_C} \left(\frac{t}{\tau_C} \right)^{\alpha'-1} \mathrm{E}_{\alpha', \alpha'} \left(-\left(\frac{t}{\tau_C} \right)^{\alpha'} \right) \sim
  \begin{cases}
	 \frac{1}{\tau \mathrm{\Gamma}(\alpha')}  \left(\frac{t}{\tau_C}\right)^{\alpha'-1}, \enspace & t \ll \tau_C \\ 
	 \frac{1}{\tau \mathrm{\Gamma}(-\alpha')}  \left(\frac{t}{\tau_C}\right)^{-\alpha'-1}, \enspace & t \gg \tau_C,
\end{cases}
\label{eq:Mittag-Leffler2Asymptote}
\end{equation}
where $\mathrm{E}_{\alpha,\beta}$ is the two-parameter Mittag-Leffler function. 
The small time approximation is the Curie--von Schweidler law from 1889/1907. 

Likewise, in \cite{holm2020time} it is shown that the charge step response is proportional to the relaxation function, $\Psi(t)$, of \cite{garrappa2016models}. The relaxation function is the inverse Fourier transform of $\left( 1-\varepsilon'(\omega)\right)/{\J \omega}$ and is for the Cole--Cole model:
\begin{equation}
 \Psi(t) = \mathrm{E}_{\alpha', 1} \left(-\left(\frac{t}{\tau_C}\right)^{\alpha'} \right) \sim
 \begin{cases}
	 \exp{[\frac{-(t/\tau_C)^{\alpha'}}{\mathrm{\Gamma}(\alpha'+1)}]}, \enspace & t \ll \tau_C \\
	 \frac{1}{\mathrm{\Gamma}(1-\alpha')} \left(\frac{t}{\tau_C} \right)^{-\alpha'}, \enspace & t \gg \tau_C.
\end{cases}
\label{eq:Mittag-LefflerAsymptote}
\end{equation}
The small-time approximation is a stretched exponential which corresponds to the Kohlrausch law \cite{holm2020time} from 1854. 
This demonstrates the close relationship between the CPE, the standard dielectric models, and the classical current and charge responses.

\subsubsection{Inductive CPE}

An inductive CPE is defined by extension of the results for the capacitive CPE \cite{holm2021simple}:
\begin{equation}
{Z}(\omega) = (\I \omega)^{\alpha} L_{\alpha},  \enspace 0<\alpha \le 1.
\label{eq:fractionalL}
\end{equation}
This can be considered to be a capacitive CPE where $\alpha' = -\alpha$ and it will have a similar temporal response as that of \eqref{eq:Mittag-Leffler2Asymptote} with voltage and current changing roles, i.e.~voltage proportional to  $t^{-\alpha-1}$ for small temporal arguments. The similarity with the temporal response of the linearly varying capacitance in \eqref{eq:ApproxLinearlyVaryingC} is striking, and indicates one way of implementing a CPE as noted in \cite{holm2021simple}.

\section{Method}

The results reported here were found during an attempt to make a circuit with the same voltage response as the inductive CPE, exploiting the just mentioned similarity in responses. In our first attempt at circuit realization we used the circuit of Fig.~2 of \cite{senani1992simple} consisting of two op-amps and a  voltage-controlled resistor  configured to yield a voltage-controlled capacitor. The voltage-controlled resistor was implemented by a JFET operating in the non-saturated or triode region. 
The circuit was not so successful due to the restricted range of variation of the JFET's resistance that was possible to achieve.

\begin{figure}[tb]
\centering
\includegraphics[scale=0.35]{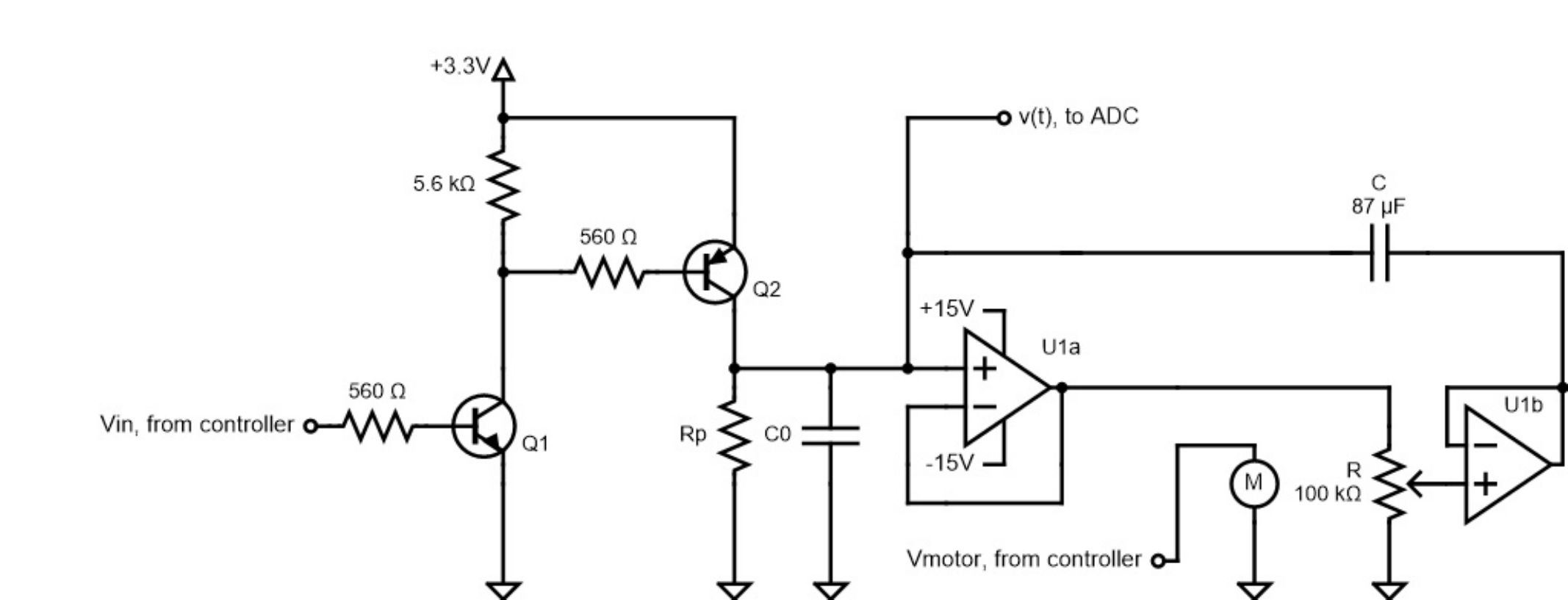}
\caption{Schematics with time-varying capacitor implemented by a time-varying resistor. Q1=BC547C, Q2=BC557B, and U1=LM358P (Drawn using \url{https://www.circuit-diagram.org/}).}
\label{fig:mekvoltdivider}
\end{figure}

To improve  performance, the JFET was substituted with a potentiometer  mechanically coupled to a stepper motor using a 3D-printed adapter. The stepper motor was controlled by a microcontroller programmed in the Arduino programming environment. In this way a precise time-varying resistor can be realized, but as it is a circuit with physically moving parts it will be much slower than the JFET circuit. The circuit diagram is shown in Fig.~\ref{fig:mekvoltdivider}.

The components used were:
\begin{itemize}
    \item Potentiometer: Bourns 91A1A-B28-B20, 240$^\circ \pm 5^\circ$,  \SI{100}{k \Omega} $\pm 20\%$, 1 Watt, linear taper, measured to be $R=\SI{109}{k\Omega}$
    \item Motor: 28BYJ-48, nominally 2048 steps per revolution
    \item Microcontroller: Adafruit Metro M0 Express
\end{itemize}
%
The capacitor  was made by connecting eight  ceramic capacitors with nominal value $\SI{10}{\micro F}$ in parallel. The total capacitance was measured with a component tester \cite{ttester} to have a value of  $C=\SI{87}{\micro F}$. This particular component tester measures the capacitance by means of the time constant for a discharge, which is similar to how the capacitor is used in our circuit.

The purpose of $Q_1$ and $Q_2$ was to set the initial voltage $v(t)=V_0$ over $R_p$ and the time-varying capacitor. As the motor starts rotating the potentiometer, transistor $Q_2$ is opened so that the monitored voltage $v(t)$ is no longer fixed, but is free to vary. The op-amps used are not very critical, since the frequency is very low. The most important parameters are a high open-loop gain, a high input impedance, and a low output impedance. The LM358 dual op-amp was used, with satisfactory results.

The measurement of $v(t)$ was performed by using a 16-bit analog to digital converter (Adafruit ADS1115) connected to a separate microcontroller and sampling at 860 Hz.

First the motor-driven potentiometer was connected to a DC source and it was verified that it actually worked as a linearly varying resistor. 
The time $\Delta t$ from zero to max resistance was $\Delta t\approx\SI{2.64}{s}$. This gives the value for the rate of change of the resistance, $\phi$:
\begin{equation}
    \phi=\od{R}{t}=\frac{\SI{109}{k\Omega}}{\SI{2.64}{s}}\approx\SI{41.3}{k\Omega/s}.
\end{equation} 
The number of steps of the stepper motor used in practice is due to the limited rotation angle of the potentiometer $2048 \cdot 240^\circ / 360^\circ = 1365$.

The  time-varying capacitor as seen between ground and the input of op-amp $U_1$, marked $v(t)$, is given by the value of the feedback capacitor $C$ and the Miller effect. The negative voltage gain of the two unity gain amplifiers, $U_1$ and $U_2$, and the potentiometer combined is:
\begin{equation}
    A_v = - \frac{R-r(t)}{R},
\end{equation}
where $r(t)$ is the resistor of the upper part of the potentiometer, $R$. The input impedance as a function of the impedance of the feedback capacitor, $Z_C$, is:
\begin{equation}
    Z_{in} = \frac{Z_C}{1+ A_v} = \frac{1}{\J \omega C} \frac{R}{r(t)}.
\end{equation}
When the capacitor $C_0$ also is taken into account, the equivalent capacitance is:
\begin{equation}
    C(t)=C_0+C\frac{r(t)}{R}=C_0+C\frac{\phi t}{R}.
\end{equation}
The coefficient of variation of the capacitor, $\theta$ of \eqref{eq:TimeVaryingC}, therefore takes the value
\begin{equation}
    \theta=\od{C(t)}{t}=C\frac{\phi}{R}=\frac{C}{\Delta t} \approx \frac{\SI{87}{\micro F}}{\SI{2.64}{s}} \approx \SI{33} {\micro F/s}.
\end{equation}
The capacitor $C_0$ was not implemented by any physical capacitor. It represents mechanical effects in the motor-potentiometer system and its value will be estimated from the measurements.

\begin{figure}[tb]
\centering
\includegraphics[scale=0.7]{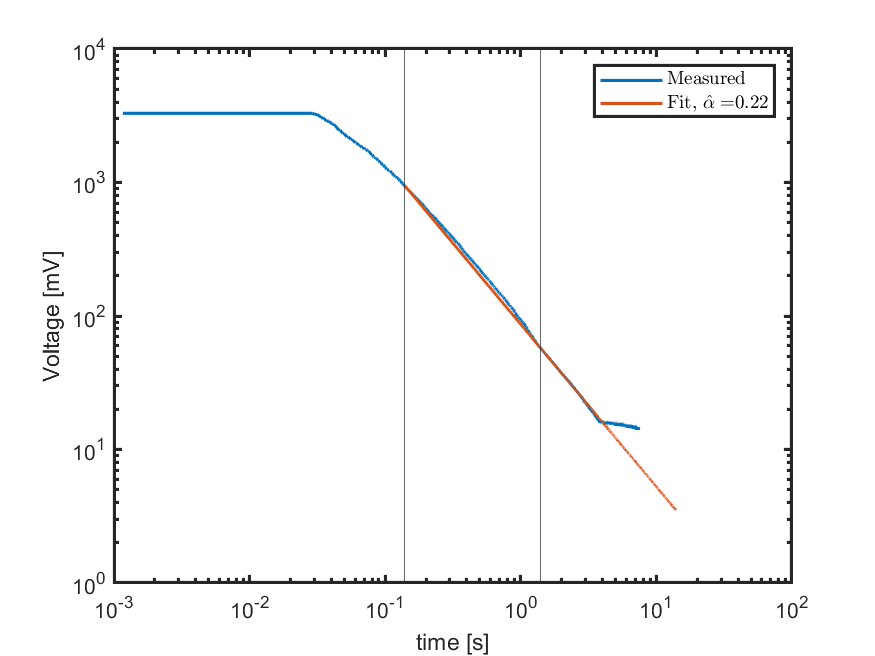}
\caption{Results with a parallel resistor of $\SI{150}{k\Omega}$}
\label{fig:R=150k}
\end{figure}

\begin{figure}[tb]
\centering
\includegraphics[scale=0.7]{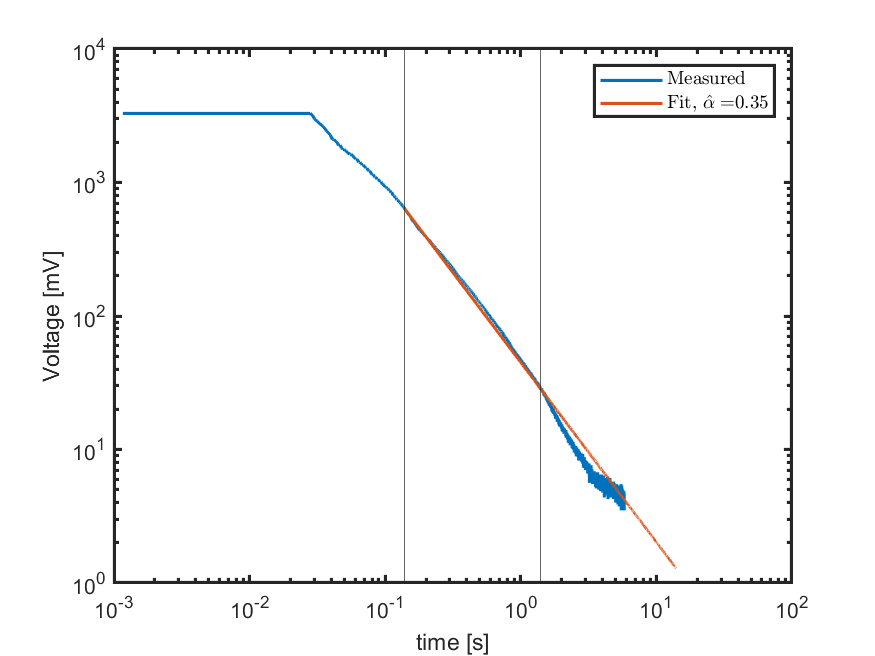}
\caption{Results with a parallel resistor of $\SI{82}{k\Omega}$}
\label{fig:R=82k}
\end{figure}

\section{Results and Discussion}
\subsection{Variable Capacitor Circuit}
\begin{figure}[tb]
\centering
\includegraphics[scale=0.7]{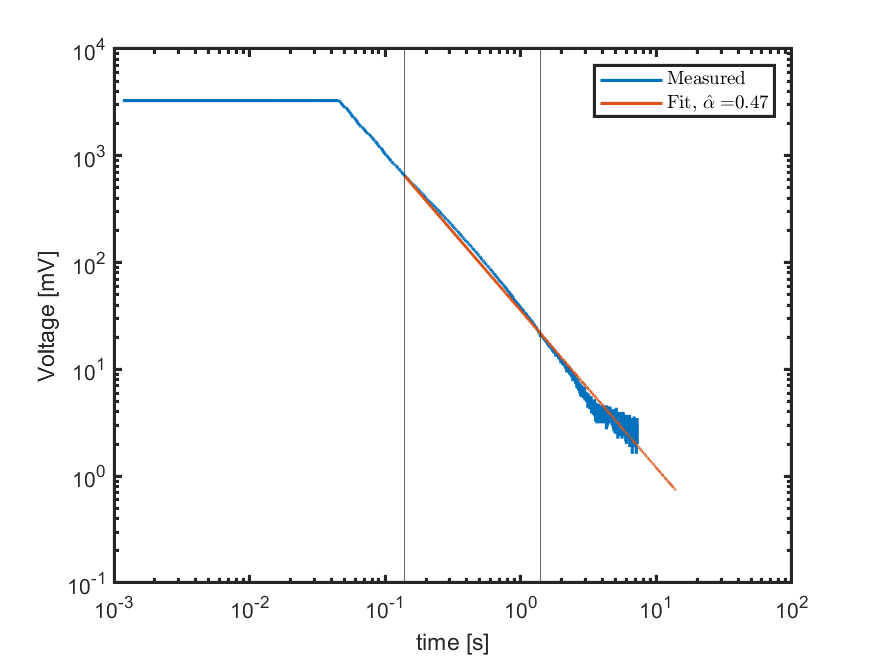}
\caption{Results with a parallel resistor of $\SI{68}{k\Omega}$}
\label{fig:R=68k}
\end{figure}

\begin{figure}[tb]
\centering
\includegraphics[scale=0.7]{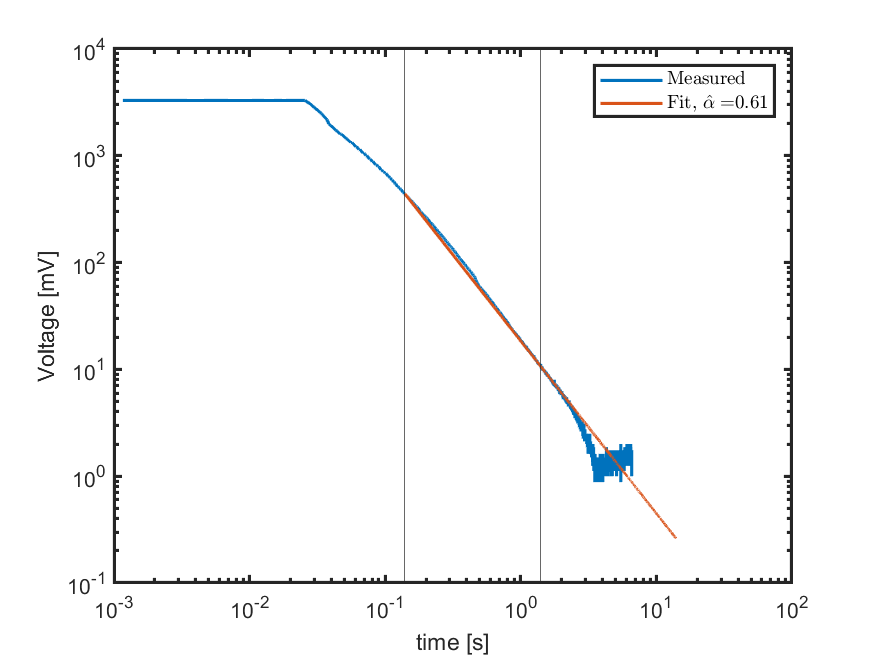}
\caption{Results with a parallel resistor of $\SI{47}{k\Omega}$}
\label{fig:R=47k}
\end{figure}

The results were obtained by measuring the voltage response to an inital input voltage for different values of the parallel resistor. The  measurements are reported in Figs.~\ref{fig:R=150k} -- \ref{fig:R=47k}. The time before the curves start falling represents the time constant, $\tau$, of \eqref{eq:LCPEimpulse2}. It is caused by mechanical slack in the motor-potentiometer system and is in the range 25--$\SI{50}{m s}$. It can therefore safely be neglected compared to the measurement time of \SI{2.64}{s} verifying that we can expect the simple power-law expression of \eqref{eq:ApproxLinearlyVaryingC} to be accurate (provided that \eqref{eq:I-TimeMultiplication} is correct). This range of values for $\tau$ corresponds to $C_0$ in the order of $\SI{1}{\micro F}$. 

The expected $\alpha$-values were calculated from \eqref{eq:LCPEimpulse2} as 
\begin{equation}
    \alpha = \frac{\Delta t}{R_pC}=\frac{\SI{2.64}{s}}{R_p\cdot\SI{87}{\mu F}},
\end{equation}
and $R_p$ and the expected value for $\alpha$ are shown in the first columns of Table \ref{tab:my-table}.

\begin{table}[t]
\begin{center}
\begin{tabular}{|r|c|c|c|}
\hline
 $R_p$  & $\alpha$    &$\hat{\alpha}$ & Figure\\\hline
 150k   & 0.20              & 0.22          & \ref{fig:R=150k}  \\\hline
  82k   & 0.37              & 0.35          & \ref{fig:R=82k}   \\\hline
  68k   & 0.45              & 0.47          & \ref{fig:R=68k}   \\\hline
  47k   & 0.65              & 0.61          & \ref{fig:R=47k}   \\\hline
\end{tabular}
\caption{Table of  resistor values, $R_p$, expected value of $\alpha$, and estimated value $\hat{\alpha}$.}
\label{tab:my-table}
\end{center}
\end{table}

Beyond a time of $\Delta t \approx 2.64$ s, the curves deviate from a power law, as the potentiometer has stopped moving and the circuit behaves like a simple RC-circuit with an exponential fall-off.

The estimate of order is shown in these figures and in the third column of Table \ref{tab:my-table}. The estimate was found by picking measured values over one decade in the range $\tau < t < \Delta t$. The vertical lines at $t_1=139.5$ ms and $t_2=10t_1$ show the values used. The estimate is:
\begin{equation}
\hat{\alpha} = -\frac{\ln{v(t_2)} - \ln{v(t_1)}}{\ln{t_2} - \ln{t_1}} - 1.
\end{equation}

As can be seen from  the figures, the relaxation  follows a power law for about two decades of time. This is adequate for concluding that the response is the power law of \eqref{eq:ApproxLinearlyVaryingC} and not the exponential decay of \eqref{eq:Exponential}. Comparing columns 2 and 3 in  Table \ref{tab:my-table}, there is a difference in the order of 0.01--0.04 which indicates that there is also a good correlation between the theoretic $\alpha$-value expected from the component values, and the estimated value from the measured data.

\section{Conclusion}
 The voltage response of the parallel combination of a resistor and a linearly increasing capacitor was measured to follow a power law over about two decades of time and the exponent matched the expected value from the theoretical model. This is confirmation that the two-term time-domain multiplication definition of current in a capacitor of \eqref{eq:I-TimeMultiplication} is correct for this circuit in accordance with what is stated in  \cite{holm2021simple, jeltsema2022time, ortigueira2023new}. This is in contrast to the Constant Phase Element. When interpreted as a time- and frequency-varying capacitance, as is usually done, it follows the time domain convolution of \eqref{eq:I-TimeConvolution} in agreement with \cite{allagui2021revisiting, allagui2023tikhonov}. This demonstrates that a CPE is a fundamentally different component from a time- and frequency-varying capacitance.

 \bibliographystyle{elsarticle-num} 
 \bibliography{timeVarying}





\end{document}